# A Dual Beam Hα Doppler System to Acquire, Analyse and Anticipate Solar Eruptive Events Directed towards Earth

Anand D. Joshi[1], Shibu K. Mathew[1], Nandita Srivastava[1], Sara F. Martin[2], Sudhir Kumar Gupta[1]

[1]Udaipur Solar Observatory, Physical Research Laboratory, Udaipur, India.
[2]Helio Research, La Crescenta, USA.

A new instrument with a dual-beam Hα Doppler system is being developed at the Udaipur Solar Observatory (USO) in order to improve the quality and quantity of data on quiet, activated and erupting filaments and prominences on the Sun, especially those associated with geo-effective coronal mass ejections. These data can be potentially used to construct three-dimensional topology of erupting filaments as they leave the surface of the Sun and can be compared with multi-wavelength data obtained from space missions such as STEREO, SOHO, and Hinode. The characterization of various optical components for the instrument is being carried out, and some preliminary results are described in the paper.

## 1 Introduction

It is found that a majority of coronal mass ejections (CMEs) are associated with erupting filaments, and many of them are directed towards Earth. Such CMEs can cause geomagnetic storms, when magnetic field carried by CMEs reconnects with Earth's magnetic field. Webb et al. (1976) observed enhancements in X-ray images following disappearances of (quiescent) filaments seen in Hα. The X-ray enhancements occurred at the same location as the Hα filaments. Munro et al. (1979) have found that more than 70% of CMEs had associated eruptive prominences or filament disappearances. Recently, Gopalswamy et al. (2003) analysed all CMEs from January 1996 up to December 2001 observed by the Large Angle Spectrometric Coronagraph (LASCO) on board Solar and Heliospheric Observatory (SOHO), and compared them with prominence events seen from the Nobeyama Radioheliograph (NoRH). Out of the 186 prominence events covered by both LASCO and NoRH, they found that 134 (72%) had near simultaneous (~ 30 minutes) CMEs associated with them. Srivastava & Venkatakrishnan (2004) have found that erupting filaments, in particular, close to central meridian and having low-latitudes, are important, as they are most likely to give rise to severe geomagnetic storms at Earth. Moreover, intensities of geomagnetic storms are correlated with initial speeds of CMEs (Srivastava & Venkatakrishnan, 2002). Although CMEs are only observed above the limb, the cited close relationships between erupting filaments and CMEs is important. If there is to be progress in predicting when an Earth-directed CME will intersect the Earth's magnetosphere and cause a geomagnetic storm, it is imperative to record and study Doppler signals from filaments, which are likely to erupt toward Earth with an associated CME.

A unique instrument used for monitoring filament activations and for recording line-of-sight velocities of erupting filaments has been developed and operated at Helio Research. This instrument is identified as a "dual beam Hα Doppler system". In order to increase the hours of solar monitoring and recording of high quality Hα Doppler data per 24 hour day, it was decided to build an identical instrument at USO. The data from this instrument can be compared with other space-based instruments for continuous monitoring of the solar activity.

## 2 Overview of instrument

The main components of the proposed dual beam Hα Doppler system are a tilt-tunable pre-filter with 1 Å passband and a voltage tunable lithium niobate ($LiNbO_3$) Fabry-Perot etalon (FP) with 0.1 Å passband. A schematic diagram of the set-up is given in figure 1. It is found that $LiNbO_3$ etalons can provide high quality Doppler images and Dopplergrams in Hα (Rust et al., 1987; Mathew et al., 1998). The 1 Å pre-filter will act as a stand-alone filter for monitoring activity of quiescent prominences in one beam of the instrument, henceforth referred to as 'mode a', and also as a blocking filter in front of the 0.1 Å LiNbO$_3$ FP, henceforth referred to as 'mode b'. By tuning the FP across the Hα line, mode b will give line-of-sight Doppler velocities of erupting filaments with better velocity discrimination but with lower contrast and velocity range than the 1 Å filter.

The 1 Å pre-filter is designed specifically for the purpose of this instrument at Helio Research. It consists of two 1.4 Å single-period interference filters, IF1 and IF2 in figure 1. Sandwiched between them are a linear polarizer (LP) and a quarter wave plate (QWP), which eliminate internal reflections between the two IFs (IF1 and IF2).

The substrate of FP is made of LiNbO$_3$ crystal which is birefringent and electro-optic. The FP can be tuned across the Hα line profile by applying suitable voltages. Figure 2 gives the method of determining Doppler velocity. For a Doppler-shifted Hα line, the difference in intensities measured at $\lambda_0+\Delta\lambda$ and $\lambda_0-\Delta\lambda$, where $\lambda_0$ is Hα line centre (6562.8 Å), is directly proportional to the velocity (v).

$$I_{blue} - I_{red} \propto \lambda - \lambda_0$$
$$\text{and } \frac{\lambda - \lambda_0}{\lambda_0} = \frac{v}{c}$$

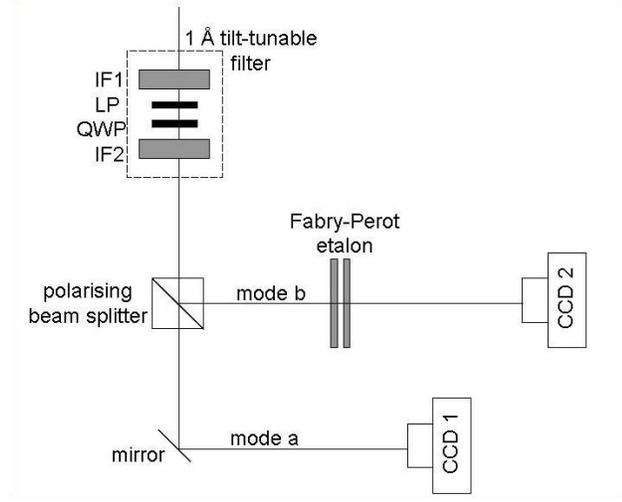

Fig. 1 : Schematic of proposed set-up at USO

The identical instrument operating at Helio Research has obtained good quality images such as the four filaments in Figure 3 These images are taken through the 1 Å pre-filter, corresponding to 'mode a' for our instrument. However, of the four images, only one has distinctly higher contrast (panel b). This is because it was an activated filament, that erupted within 1 hour as seen in figure 4. This shows that the 1 Å pre-filter can be used as a primary monitor for the pre-eruptive status of quiescent filaments, and also can provide data to measure line-of-sight velocities of erupting filaments. However, if one's goal is to measure the velocities with better resolution, the $LiNbO_3$ the FP should be used to take narrow band images simultaneously.

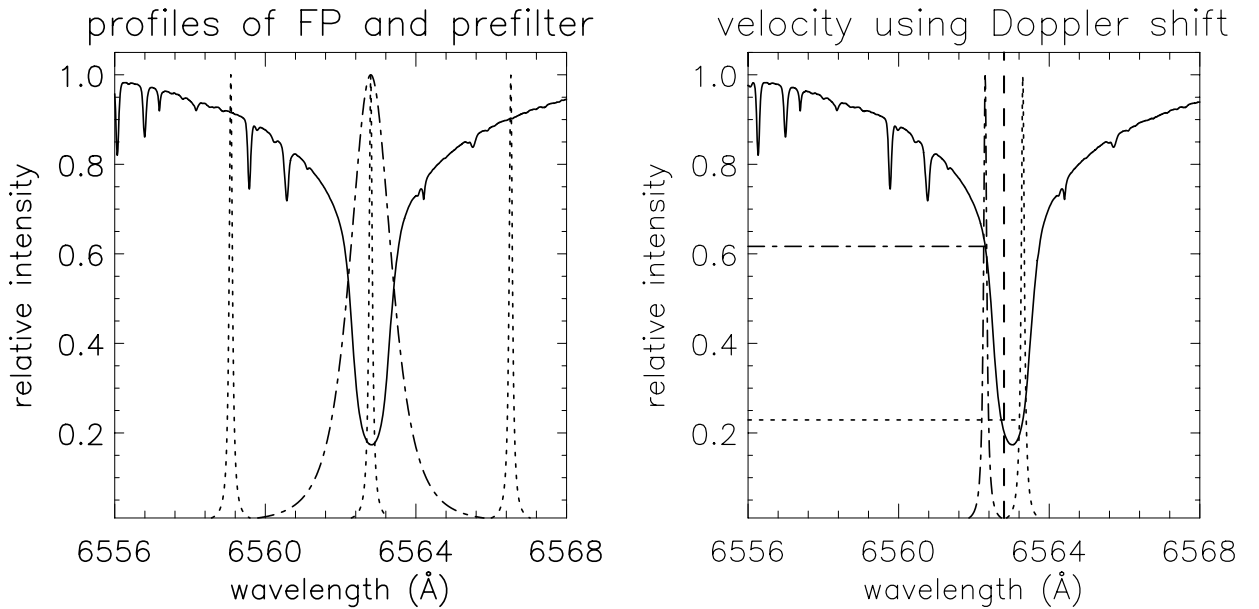

Fig. 2 : LEFT panel: solid: Hα line profile, dash-dotted: pre-filter transmission profile, dotted: FP channel spectra
RIGHT panel: solid: shifted Hα line profile, dashed: Hα line centre, dash-dotted: Hα – 0.5 Å, dotted: Hα + 0.5 Å

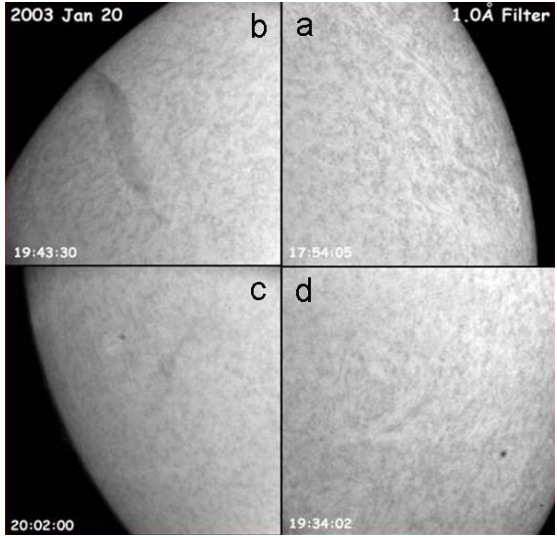

Fig. 3: Images taken directly through the 1 Å filter show 3 quiet (panels a, c and d) and 1 activated filament (panel b).

## 3 Calibration set-up

The proposed instrument being developed at USO uses a Littrow spectrograph for calibration, testing and characterization of the FP and pre-filter. A Coudé refractor telescope with primary lens of focal length 2.5 m feeds sunlight to the slit of the spectrograph. A collimating lens with focal length 0.80 m directs the light onto a diffraction grating having 1200 lines $mm^{-1}$. The grating is blazed at 22° to give maximum efficiency in the $2^{nd}$ order. It is mounted on a rotating stage so that the required part of spectrum can be selected. The linear dispersion obtained by this set-up at the plane of CCD camera is about 29 mÅ per pixel, and covers a spectral range of 17 Å around Hα.

A high voltage power supply procured from Applied Kilovolts is used to tune the FP. The power supply is capable of producing output from –5 kV to +5 kV, but will be used in the range of –3.0 kV to +3.0 kV This limit prevents the possible damage of the Fabry-Perot due to very high voltage. The power supply is controlled by a low voltage input ranging between –10 V and +10 V. A 16-bit digital-to-analog converter and the required software is built at USO for this purpose, which gives a voltage resolution of

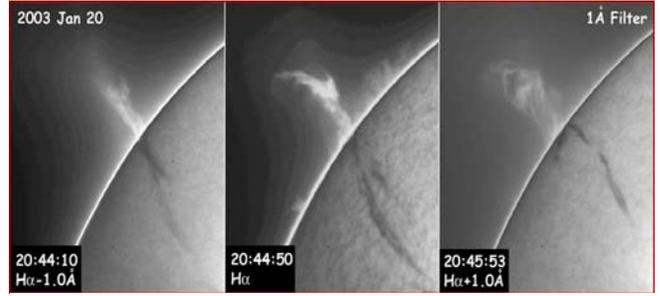

Fig. 4: Subsequent eruption of the activated filament (seen in panel b in figure 3) observed in three wavelength positions obtained by tilting the 1 Å pre-filter.

1 V. High voltage is applied in steps of 10 V $ms^{-1}$ to ensure the safe operation of the FP.

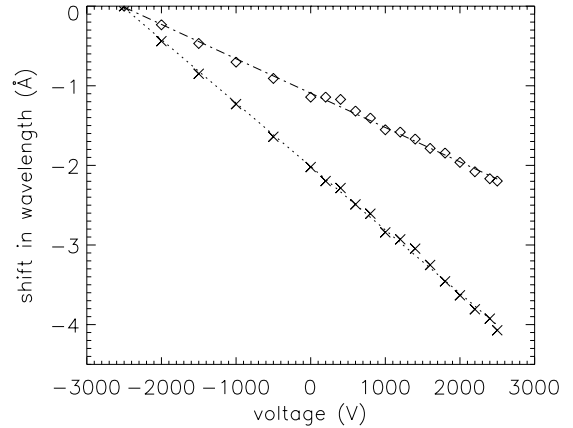

Fig. 5: Shift in wavelength (crosses for o channel and diamonds for e channel) plotted against applied voltage. Dotted and dash-dotted lines show linear fits for o and e channels respectively.

To check the voltage calibration, we applied voltage to the FP from -2.5 kV to +2.5 kV, and checked for the shift in wavelength. The plot for the same is given in figure 5 while the details of the ordinary (o) and extraordinary (e) channels are given in table 1. We can see that the voltage tunability for ordinary channel is close to twice that of the extraordinary channel., Hence, it is of greater advantage for us to use the o channel. Half FSR voltage is the voltage needed to shift the channel by one half of the free spectral range (FSR). Thus for the o channel, we need to apply 2210 V at most to bring the passband to the desired wavelength.

|  | o channel | e channel |
|---|---|---|
| FSR (Å) | 3.60 | 3.74 |
| voltage tunability (mÅ/V) | – 0.80 | – 0.43 |
| half FSR voltage (V) | 2210 | 4256 |

Table 1: Properties for ordinary and extraordinary channels

Due to its sensitivity to temperature, the FP is enclosed in a temperature controlled oven maintained at 43 °C, which is the maximum ambient temperature during summer in Udaipur. The oven, which is built at USO, is accurate up to 0.0625 °C. A 12-bit digital temperature sensor, DS620 from Dallas Semiconductors, is used for measuring the temperature. The oven is designed such that in case the temperature inside exceeds 47 °C, power to the heating element is cut, and is restored only after the temperature falls below 45 °C. The temperature of oven was raised up to 43 °C, from the ambient temperature, and the growth curve obtained is plotted in figure 6. We can see the temperature being stable within 0.0625 °C of the set value. In the background, for each value of time, a part of the corresponding channel spectrum (5 pixels wide with spectral axis along the vertical) captured by the CCD is shown. Three maxima each of the ordinary (upper) and extraordinary (lower) channels are seen stacked in time. As temperature increases, the o and e channels are seen to merge because their responses to change in temperature are different.

**4 Summary**

The dual beam Hα Doppler instrument being built at Udaipur Solar Observatory will be fully operational when the set-up and calibration for both modes a and b is complete. At present, we are working with the LiNbO$_3$ FP beam alone. We plan to deploy the 1 Å pre-filter once temperature and voltage calibrations are completed on the spectrograph. Later we will be using the set-up for imaging Sun using a CCD camera with a 4k x 4k Kodak chip, bought from Finger Lakes Instrumentation (FLI). This instrument would enable us to monitor activations of filaments as well as measure line-of-sight velocities of erupting ones.

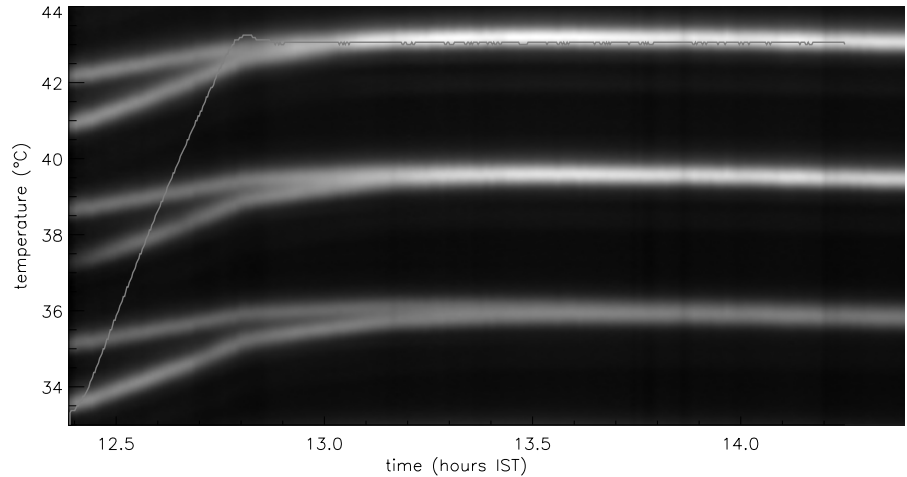

Fig. 6: Rise in temperature of Fabry-Perot etalon plotted against time. Image in background shows part of channel spectrum captured.